\documentclass[sigconf,natbib=true,anonymous=false]{acmart}
\usepackage{pgfplots}
\usepackage{booktabs}
\usepackage{caption}
\AtBeginDocument{%
  }

\copyrightyear{2025}
\acmYear{2025}
\setcopyright{cc}
\setcctype{by-sa}
\acmConference[SIGIR '25]{Proceedings of the 48th International ACM SIGIR
Conference on Research and Development in Information Retrieval}{July 13--18,
2025}{Padua, Italy}
\acmBooktitle{Proceedings of the 48th International ACM SIGIR Conference on
Research and Development in Information Retrieval (SIGIR '25), July 13--18,
2025, Padua, Italy}\acmDOI{10.1145/3726302.3730163}
\acmISBN{979-8-4007-1592-1/2025/07}

\usepackage[normalem]{ulem} 
\usepackage{tabularx}
\usepackage{booktabs}
\usepackage{arydshln}
\usepackage{multirow}
\usepackage{graphicx}
\usepackage{colortbl}
\usepackage{todonotes}
\usepackage{bbm}
\usepackage{adjustbox}
\usepackage[explicit]{titlesec}
\usepackage[moderate,leading=normal,paragraphs=tight]{savetrees}

\titlespacing{\paragraph}{%
  0em}{
  0\baselineskip}{
 0\baselineskip}%

\begin{document}

\title[An Alternative to FLOPS Regularization to Effectively Productionize SPLADE-Doc]{An Alternative to FLOPS Regularization \\ to Effectively Productionize SPLADE-Doc}


\author{Aldo Porco}
\authornote{These authors contributed equally to this work.}
\affiliation{
  \institution{Bloomberg} 
  \country{New York, United States}
}
\email{aporco2@bloomberg.net}

\author{Dhruv Mehra}
\authornotemark[1]
\affiliation{
  \institution{Bloomberg} 
  \country{New York, United States}
}
\email{dmehra19@bloomberg.net}

\author{Igor Malioutov}
\authornotemark[1]
\affiliation{
  \institution{Bloomberg} 
  \country{New York, United States}
}
\email{imalioutov@bloomberg.net}

\author{Karthik Radhakrishnan}
\authornotemark[1]
\affiliation{
  \institution{Bloomberg} 
  \country{New York, United States}
}
\email{kradhakris10@bloomberg.net}

\author{Moniba Keymanesh}
\authornotemark[1]
\affiliation{
  \institution{Bloomberg} 
  \country{New York, United States}
}
\email{mkeymanesh1@bloomberg.net}

\author{Daniel Preo\c{t}iuc-Pietro}
\affiliation{
  \institution{Bloomberg} 
  \country{New York, United States}
}
\email{dpreotiucpie@bloomberg.net}

\author{Sean MacAvaney}
\affiliation{
  \institution{University of Glasgow}
  \country{United Kingdom}
}
\email{sean.macavaney@glasgow.ac.uk}

\author{Pengxiang Cheng}
\affiliation{
  \institution{Bloomberg} 
  \country{New York, United States}
}
\email{pcheng134@bloomberg.net}


\begin{abstract}
  Learned Sparse Retrieval (LSR) models encode text as weighted term vectors, which need to be sparse to leverage 
  inverted index structures during retrieval.
  SPLADE, the most popular LSR model, uses FLOPS regularization to encourage vector sparsity during training.
  However, FLOPS regularization does not ensure sparsity among terms---only within a given query or document.
  Terms with very high Document Frequencies (DFs) substantially increase latency in production retrieval engines, such as Apache Solr, due to their lengthy posting lists.
  To address the issue of high DFs, we present a new variant of FLOPS regularization: \textit{DF-FLOPS}.
  This new regularization technique penalizes the usage of high-DF terms, thereby shortening posting lists and reducing retrieval latency.
  Unlike other inference-time sparsification methods, such as stopword removal, \textit{DF-FLOPS} regularization allows for the selective inclusion of high-frequency terms in cases where the terms are truly salient.
  We find that DF-FLOPS successfully reduces the prevalence of high-DF terms and lowers retrieval latency (around 10$\times$ faster) in a production-grade engine while maintaining effectiveness both in-domain (only a 2.2-point drop in MRR@10) and cross-domain (improved performance in 12 out of 13 tasks on which we tested).
  With retrieval latencies on par with BM25, this work provides an important step towards making LSR practical for deployment in production-grade search engines.
\end{abstract}



\maketitle

\renewcommand{\shortauthors}{Aldo Porco et al.}

\vspace{-0.2cm}
\section{Introduction and Related Work}
\begin{figure}[t]
\centering
\includegraphics[width=0.75 \linewidth, , height=2.3cm]{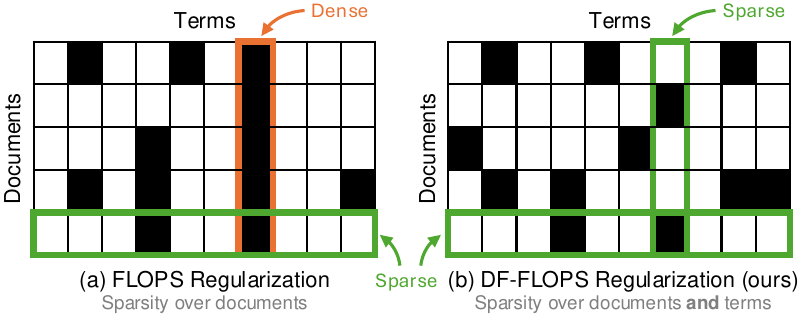}
\caption{ \small FLOPS (a) enforces sparsity within documents, but not over terms. DF-FLOPS (b) addresses this by enforcing sparsity over both documents and terms.
\vspace{-0.2cm}
}
\label{fig:overview}
\end{figure}

Learned Sparse Retrieval~(LSR)~\cite{DBLP:conf/ecir/NguyenMY23} is a prominent family of retrieval methods that includes models such as SPLADE~\cite{DBLP:conf/sigir/FormalPC21}, DeepImpact~\cite{DBLP:conf/sigir/MalliaKST21}, EPIC~\cite{DBLP:conf/sigir/MacAvaneyN0TGF20b}, UniCOIL~\cite{DBLP:journals/corr/abs-2106-14807}, and TILDE~\cite{DBLP:conf/sigir/ZhuangZ21a}. These methods aim to retrieve and rank by encoding queries and documents as vectors consisting of terms mapped to a weight (importance, salience, or impact), akin to the bag-of-words vectors used by traditional lexical retrieval models. Unlike traditional lexical models, LSR vectors are not limited to terms that appear in the original text; they are usually expanded using techniques such as language modeling expansion~\cite{DBLP:conf/sigir/MacAvaneyN0TGF20b} or external expansion~\cite{DBLP:conf/sigir/MalliaKST21}. The introduction of expansion means that the vectors do not adhere to the typical term sparsity of language without special intervention. Sparsity is critical for efficient retrieval, since the denser a vector is, the more operations will be required to score it during retrieval. A variety of techniques have been proposed to help enforce sparsity in LSR models, both at training time (e.g., FLOPS~\cite{DBLP:conf/sigir/FormalPC21} and L$_1$~\cite{DBLP:conf/ecir/NguyenHYR24} regularization), and inference time (e.g., top-k pruning~\cite{DBLP:conf/sigir/MacAvaneyN0TGF20b, lassance2023static}, score thresholding~\cite{DBLP:conf/sigir/LassanceLDCT23}, and stopword removal~\cite{DBLP:conf/sigir/LassanceC22}).

Several advances have made LSR models more practical for use in production environments. For instance, SPLADE-Doc~\cite{spladev2} eliminates the query encoder, relying instead on the literal (binary) tokens in the query, thereby nearly eliminating the cost of preparing a query for retrieval. However, a major remaining practical challenge is the length of posting lists. Indeed, LSR models tend to learn to generate some terms in nearly every vector~\cite{DBLP:journals/corr/abs-2110-11540}, which results in high Document Frequencies (DFs) and lengthy posting lists. 
Past work attempts to address this issue with static pruning~\cite{lassance2023static} and dynamic pruning with distillation~\cite{qiao2023representation}. However, these methods do not study the effects of DF on latency. Although optimizations such as BlockMaxWAND~\cite{DBLP:conf/sigir/DingS11} allow for skipping portions of lengthy posting lists, full posting list traversal is often unavoidable in production search engines because some functionalities, such as filtering, requires accessing the entire list. Bespoke LSR engines have been proposed~\cite{DBLP:conf/sigir/BruchNRV24,DBLP:conf/cikm/BruchNRV24,DBLP:conf/sigir/MalliaST24} and are orthogonal to this research, since they also benefit from decreased DFs in the first stages.

In this work, we propose a new method (DF-FLOPS) that mitigates latency issues in LSR models by penalizing terms with high document frequencies. DF-FLOPS can be seen as a generalization of FLOPS regularization that encourages sparsity both within a given document and across terms in the vocabulary (Fig~\ref{fig:overview}).  Unlike approaches that reduce DFs at inference time (such as term-centric quantile pruning~\cite{DBLP:conf/sigir/LassanceLDCT23} or stopword removal~\cite{DBLP:conf/sigir/LassanceC22}), applying DF-FLOPS regularization during training ensures that the model avoids over-relying on high DF tokens, while still allowing stopwords to be used when appropriate (e.g., when they convey meaning, such as ``who'' when it represents the World Health Organization). An example of such cases is in Table~\ref{example}, where FLOPS fails to drop high DF tokens that are semantically irrelevant (like ``is"), while DF-FLOPS removes them and correctly retains tokens like ``who" that are crucial to the passage's semantics.

We focus on one production-oriented LSR model architecture: SPLADE-Doc~\cite{spladev2}. This architecture is appealing in a production setting because it only encodes the document through a transformer; the query vector is simply represented as a binary bag-of-tokens, which eliminates substantial query-time compute. When we apply DF-FLOPS to SPLADE-Doc, we find that it successfully reduces high DF terms and, thus, retrieval latency, while maintaining or improving retrieval effectiveness (both in-domain on MS-Marco~\cite{DBLP:conf/nips/NguyenRSGTMD16} and cross-domain on BEIR~\cite{thakur2021beir}).
When using FLOPS regularization, non-semantically salient terms are included in the document's vector, often with high weights. Meanwhile, when using DF-FLOPS, these terms are almost entirely omitted from the representation, keeping only terms that are salient to the document's content.

In summary, this paper proposes an alternative regularization strategy for LSR models, which reduces the DF in resulting vectors. This property makes LSR models more practical to run in real production systems, with retrieval latencies on par with BM25.

\definecolor{darkred}{rgb}{0.55, 0.0, 0.0}
\definecolor{darkgreen}{rgb}{0.0, 0.5, 0.0}

\begin{table}[t]
{\raggedright
\fontsize{8pt}{8pt}\selectfont
\rule{\linewidth}{0.8pt}
\vspace{-10pt}
\begin{center}
\textcolor{gray}{\textbf{Representation with FLOPS} }
\end{center}

\textcolor{blue}{\textbf{Document:}} Estimates \textcolor{darkred}{of} disease burden \textcolor{darkred}{and} \textcolor{black}{\sout{cost}} \textcolor{black}{\sout{-}} effectiveness \textcolor{black}{\sout{.}} \textcolor{darkred}{WHO} \textcolor{black}{\sout{/}} C \textcolor{black}{\sout{.}} Nelson \textcolor{black}{\sout{.}} Disease burden \textcolor{darkred}{is} \textcolor{darkred}{an} indicator \textcolor{darkred}{of} health \textcolor{black}{\sout{outcome}} \textcolor{black}{\sout{.}} disease burden \textcolor{darkred}{can} \textcolor{darkred}{\sout{be}} expressed \textcolor{darkred}{in} \textcolor{black}{\sout{many}} \textcolor{black}{\sout{ways}} \textcolor{black}{\sout{,}} \textcolor{darkred}{\sout{such}} \textcolor{darkred}{\sout{as}} \textcolor{darkred}{the} \textcolor{black}{\sout{number}} \textcolor{darkred}{of} \textcolor{black}{\sout{cases}} \textcolor{black}{\sout{(}} \textcolor{black}{\sout{e}} \textcolor{black}{\sout{.}} \textcolor{black}{\sout{g}} \textcolor{black}{\sout{.}} \textcolor{black}{\sout{incidence}} \textcolor{darkred}{\sout{or}} \textcolor{black}{\sout{prevalence}} \textcolor{black}{\sout{)}} \textcolor{black}{\sout{,}} \textcolor{black}{\sout{deaths}} \textcolor{darkred}{\sout{or}} \textcolor{black}{\sout{disability}} \textcolor{black}{\sout{-}} \textcolor{black}{\sout{adjusted}} life \textcolor{black}{\sout{years}} \textcolor{black}{\sout{lost}} \textcolor{black}{\sout{(}} \textcolor{black}{\sout{DALY}} \textcolor{black}{\sout{s}} \textcolor{black}{\sout{)}} \textcolor{black}{\sout{associated}} \textcolor{darkred}{\sout{with}} \textcolor{darkred}{a} \textcolor{black}{\sout{given}} \textcolor{black}{\sout{condition}} \textcolor{black}{\sout{.}} Information \textcolor{darkred}{\sout{on}} \textcolor{darkred}{the} \textcolor{black}{\sout{reported}} \textcolor{black}{\sout{incidence}} \textcolor{darkred}{of} \textcolor{black}{\sout{vaccine}} \textcolor{black}{\sout{-}} \textcolor{black}{\sout{prevent}} \textcolor{black}{\sout{able}} diseases \textcolor{darkred}{is} \textcolor{black}{\sout{provided}} \textcolor{darkred}{to} \textcolor{darkred}{WHO} \textcolor{black}{\sout{.}} 

{
\looseness=0
\textcolor{blue}{\textbf{Expansions: }} 
\textcolor{darkred}{('what', 6.18)}, \textcolor{darkred}{('is', 5.5)}, \textcolor{darkred}{('of', 5.1)}, \textcolor{darkred}{('the', 4.87)}, \textcolor{darkred}{('a', 4.83)}, \textcolor{darkred}{('who', 4.81)}, \textcolor{darkred}{('does', 4.66)}, \textcolor{darkred}{('in', 4.61)}, \textcolor{darkred}{('are', 4.4)}, \textcolor{darkred}{('an', 4.39)}, \textcolor{darkred}{('for', 4.19)}, \textcolor{darkred}{('how', 4.13)}, \textcolor{darkgreen}{('?', 4.02)}, \textcolor{darkgreen}{('disease', 3.89)}, \textcolor{darkgreen}{('burden', 3.72)}, \textcolor{darkgreen}{('diseases', 3.54)}, \textcolor{darkgreen}{('definition', 3.51)}, \textcolor{darkgreen}{('health', 3.49)}, \textcolor{darkred}{('and', 3.41)}, \textcolor{darkgreen}{('mean', 3.4)}
}

\begin{center}
\textcolor{gray}{\textbf{Representation with DF-FLOPS}}
\end{center}

\textcolor{blue}{\textbf{Document:}} 
Estimates \textcolor{darkred}{\sout{of}} disease burden \textcolor{darkred}{\sout{and}} cost \textcolor{black}{\sout{-}} effectiveness \textcolor{black}{\sout{.}} \textcolor{darkred}{WHO} \textcolor{black}{\sout{/}} \textcolor{black}{\sout{C}} \textcolor{black}{\sout{.}} Nelson \textcolor{black}{\sout{.}} Disease burden \textcolor{darkred}{\sout{is}} \textcolor{darkred}{\sout{an}} indicator \textcolor{darkred}{\sout{of}} health outcome \textcolor{black}{\sout{.}} disease burden \textcolor{darkred}{\sout{can}} \textcolor{darkred}{\sout{be}} expressed \textcolor{darkred}{\sout{in}} \textcolor{black}{\sout{many}} \textcolor{black}{\sout{ways}} \textcolor{black}{\sout{,}} \textcolor{darkred}{\sout{such}} \textcolor{darkred}{\sout{as}} \textcolor{darkred}{\sout{the}} \textcolor{black}{\sout{number}} \textcolor{darkred}{\sout{of}} cases \textcolor{black}{\sout{(}} \textcolor{black}{\sout{e}} \textcolor{black}{\sout{.}} \textcolor{black}{\sout{g}} \textcolor{black}{\sout{.}} \textcolor{black}{\sout{incidence}} \textcolor{darkred}{\sout{or}} prevalence \textcolor{black}{\sout{)}} \textcolor{black}{\sout{,}} deaths \textcolor{darkred}{\sout{or}} disability \textcolor{black}{\sout{-}} \textcolor{black}{\sout{adjusted}} \textcolor{black}{\sout{life}} \textcolor{black}{\sout{years}} \textcolor{black}{\sout{lost}} \textcolor{black}{\sout{(}} DALY \textcolor{black}{\sout{s}} \textcolor{black}{\sout{)}} \textcolor{black}{\sout{associated}} \textcolor{darkred}{\sout{with}} \textcolor{darkred}{\sout{a}} \textcolor{black}{\sout{given}} \textcolor{black}{\sout{condition}} \textcolor{black}{\sout{.}} \textcolor{black}{\sout{Information}} \textcolor{darkred}{\sout{on}} \textcolor{darkred}{\sout{the}} \textcolor{black}{\sout{reported}} \textcolor{black}{\sout{incidence}} \textcolor{darkred}{\sout{of}} vaccine \textcolor{black}{\sout{-}} \textcolor{black}{\sout{prevent}} able diseases \textcolor{darkred}{\sout{is}} \textcolor{black}{\sout{provided}} \textcolor{darkred}{\sout{to}} \textcolor{darkred}{WHO} \textcolor{black}{\sout{.}} 

{
\looseness=0
\textcolor{blue}{\textbf{Expansions: }} 
 \textcolor{darkgreen}{('burden', 3.59)}, \textcolor{darkgreen}{('disease', 3.44)}, \textcolor{darkgreen}{('estimates', 3.21)}, \textcolor{darkgreen}{('effectiveness', 3.12)}, \textcolor{darkgreen}{('diseases', 3.07)}, \textcolor{darkgreen}{('estimate', 2.98)}, \textcolor{darkgreen}{('indicator', 2.93)}, \textcolor{darkgreen}{('indicators', 2.82)}, \textcolor{darkgreen}{('nelson', 2.8)}, \textcolor{darkgreen}{('estimation', 2.67)}, \textcolor{darkgreen}{('load', 2.62)}, \textcolor{darkgreen}{('health', 2.55)}, \textcolor{darkgreen}{('vaccine', 2.53)}, \textcolor{darkgreen}{('illness', 2.52)}, \textcolor{darkgreen}{('DALY', 2.52)}, \textcolor{darkgreen}{('estimated', 2.51)}, {\setlength{\fboxsep}{0pt}\textcolor{darkred}{\colorbox{yellow}{('who', 2.44)}}}, \textcolor{darkgreen}{('weight', 2.43)}, \textcolor{darkgreen}{('cost', 2.42)}, \textcolor{darkgreen}{('effective', 2.36)}
}
\vspace{-2pt}
\rule{\linewidth}{0.8pt}

\caption{\small FLOPS and DF-FLOPS representations for a passage. 
\protect\footnotemark
}
}
\vspace{-0.5cm}

\label{example}
\end{table}

\footnotetext{Tokens are color-coded: \textcolor{darkred}{red} for stopwords, \textcolor{darkgreen}{green} for content words, and struck through (\sout{token}) if omitted by the model. Only the top 20 most weighted words in the representation—either kept in the document or in the expansions—are shown for brevity.}

\vspace{-0.2cm}
\section{FLOPS Conditioned on Document Frequency}
$FLOPS$ regularization~\cite{DBLP:conf/iclr/PariaYYXRP20} is used in SPLADE to reduce the number of non-zero elements in a document's representation, thereby reducing the number of floating point operations needed to score a document~\cite{DBLP:conf/sigir/FormalPC21}. It acts as an additional loss component that pushes the values in each representation to zero. FLOPS has been shown to perform better than other regularizers such as L1 \cite{spladev2}. Using a variation of the notation from \citet{DBLP:conf/sigir/FormalPC21}, FLOPS regularization can be formulated as follows:
\begin{small}
\begin{equation}
\ell_{FLOPS} = \sum_{t \in V} \Bigg( \frac{1}{N} \sum_{i=1}^{N} r_{i,t} \Bigg)^2
\end{equation}
\end{small}\ignorespaces
where $V$ is a set containing all terms in the lexicon, $N$ is the number of vectors in the batch, and $r_{i,t}$ is the weight of term $t$ of the $i^\text{th}$ vector in the batch (e.g., the document's representation).

The original FLOPS paper~\cite{DBLP:conf/iclr/PariaYYXRP20} theoretically demonstrated that each term has equal likelihood when it is minimized. However, in practice, we have found this not to be the case when it is applied to LSR models (Section~\ref{sec:results})\footnote{We suspect this is due to the inductive bias of the model architecture.}. This results in high latency in production environments due to terms with high DF. DF-FLOPS regularization aims to address this limitation by scaling the loss of each term by a weight $w_t$ depending on the term's document frequency. The formulation is as follows:
\begin{small}
\begin{align}
\ell_{DF-FLOPS} = \sum_{t \in V} \Bigg(  \frac{w_t}{N} \sum_{i = 1}^{N} r_{i,t} \Bigg)^2 \;,\;\;\; \text{where} \;\; 
w_t = activ\bigg(\frac{DF_t}{|C|}\bigg)
\label{eq:main}
\end{align}
\end{small}\ignorespaces
where $DF_t$ is an approximation of the document frequency of term $t$ (i.e., the number of documents that have a non-zero weight for $t$), $|C|$ is the size of the corpus over which $DF$ was approximated, and $activ(\cdot)$ is a non-linear activation function over the $DF_t\;/\;|C|$ ratio. Details on how DF can be approximated during training and the behavior of the activation function are covered in the remainder of this section. Note that FLOPS regularization can be seen as a special case of DF-FLOPS for which $w_t=1, \;\forall\;t\in V$.

\textbf{Approximating DF:} Document Frequencies are corpus-wide statistics, so computing them for a model at every training step is computationally prohibitive. Instead, we propose periodically approximating them based on the query and document vectors computed during the existing validation stage (every 100 training steps in our configuration). In the first training stage, we set all $w_t=1$, in line with the existing FLOPS regularizer. This strategy achieves reasonably up-to-date DFs while avoiding extra GPU computation. We recognize that other strategies could be used to produce DF estimates (e.g., running averages during training), but we leave these as possible explorations for future work.

\textbf{Activation and Tuning:} We propose using the following variation of the Generalized Logistic Function~\cite{richards1959flexible} as $activ(\cdot)$:
\begin{small}
\begin{align}
activ(x; \alpha, \beta) &= \dfrac{1}{1 + (x^{log_{\alpha}2} - 1)^{\beta}} 
\end{align}
\end{small}\ignorespaces
This formulation allows for calibration of FLOPS penalties using two parameters: $\alpha$ and $\beta$. $\alpha$ acts as an approximate cutoff penalty, with $DF_t\;/\;|C|>\alpha$ generally being highly penalized and those below weakly penalized. Meanwhile, $\beta$ controls the steepness of the penalty curve, so tokens whose frequency crosses $\alpha$ will receive a higher penalty with $\beta=10$ than with $\beta=1$. Although we feel that this formulation gives the model trainer good control over the behavior, we recognize that alternative activation functions exist, and leave their exploration for future work.

\vspace{-0.1cm}
\section{Experimental Setup}

The following section describes our dataset processing, training hyperparameters, and the latency benchmarking setup.

\textbf{Datasets:} All models are trained on the MS-Marco passage ranking dataset~\cite{DBLP:conf/nips/NguyenRSGTMD16}, which comprises of around 500K training queries and 8.8M passages. For validation (and for our periodic estimations of DF), we use a sample of the training set consisting of 50K \textit{query-passage} pairs, computed every 100 training steps. For testing, we use the MS-Marco dev set, TREC DL 2019~\cite{DBLP:journals/corr/abs-2003-07820} and 2020~\cite{craswell2020overview} evaluation set, and BEIR. We use standard measures of effectiveness, including nDCG@10, MRR@10, and Recall@1000.

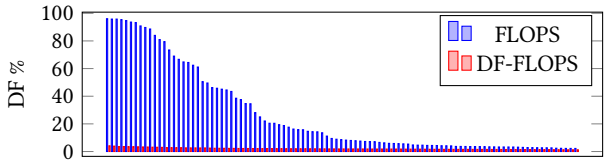
\begin{figure}[t!]
    \centering
    \pgfplotsset{height=3.5cm,width=8.5cm, compat=1.9}
\begin{tikzpicture}
\begin{axis}[
	xmajorticks=false,
	ylabel=DF~\%,
	enlargelimits=0.05,
	ybar interval=0.5,
    ylabel near ticks,
    xlabel near ticks,
    grid=none
]
\addplot 
	coordinates { (0, 96.06) (1, 95.8) (2, 95.8) (3, 95.41) (4, 94.72) (5, 93.63) (6, 93.28) (7, 90.84) (8, 89.73) (9, 88.62) (10, 84.01) (11, 80.99) (12, 79.48) (13, 73.46) (14, 69.02) (15, 66.7) (16, 64.81) (17, 64.42) (18, 62.42) (19, 61.12) (20, 50.55) (21, 49.57) (22, 46.26) (23, 45.78) (24, 45.11) (25, 44.53) (26, 43.46) (27, 38.55) (28, 37.55) (29, 34.71) (30, 34.55) (31, 28.15) (32, 25.06) (33, 22.1) (34, 20.46) (35, 20.43) (36, 19.42) (37, 18.81) (38, 17.62) (39, 16.26) (40, 15.84) (41, 15.75) (42, 14.64) (43, 14.35) (44, 14.26) (45, 13.57) (46, 11.16) (47, 9.46) (48, 8.91) (49, 8.6) (50, 8.26) (51, 8.01) (52, 7.82) (53, 7.56) (54, 7.28) (55, 7.18) (56, 7.12) (57, 6.6) (58, 6.29) (59, 5.92) (60, 5.83) (61, 5.73) (62, 5.56) (63, 5.39) (64, 4.76) (65, 4.64) (66, 4.61) (67, 4.44) (68, 4.4) (69, 4.35) (70, 4.19) (71, 4.16) (72, 4.09) (73, 3.7) (74, 3.7) (75, 3.66) (76, 3.61) (77, 3.59) (78, 3.58) (79, 3.54) (80, 3.42) (81, 3.38) (82, 3.31) (83, 3.31) (84, 3.3) (85, 3.28) (86, 3.16) (87, 3.01) (88, 2.89) (89, 2.84) (90, 2.77) (91, 2.71) (92, 2.71) (93, 2.66) (94, 2.41) (95, 2.34) (96, 2.26) (97, 2.26) (98, 2.24) (99, 2.01)};
\addplot 
	coordinates { (0, 4.15) (1, 3.93) (2, 3.62) (3, 3.61) (4, 3.54) (5, 3.51) (6, 3.42) (7, 3.32) (8, 3.29) (9, 3.14) (10, 3.1) (11, 3.02) (12, 2.86) (13, 2.82) (14, 2.8) (15, 2.79) (16, 2.67) (17, 2.65) (18, 2.63) (19, 2.61) (20, 2.6) (21, 2.49) (22, 2.37) (23, 2.35) (24, 2.34) (25, 2.28) (26, 2.28) (27, 2.22) (28, 2.21) (29, 2.19) (30, 2.18) (31, 2.16) (32, 2.15) (33, 2.15) (34, 2.13) (35, 2.1) (36, 2.09) (37, 2.06) (38, 2.03) (39, 2.03) (40, 2.02) (41, 2.02) (42, 2.01) (43, 1.98) (44, 1.96) (45, 1.94) (46, 1.93) (47, 1.91) (48, 1.9) (49, 1.89) (50, 1.88) (51, 1.88) (52, 1.87) (53, 1.87) (54, 1.86) (55, 1.83) (56, 1.82) (57, 1.82) (58, 1.81) (59, 1.81) (60, 1.81) (61, 1.79) (62, 1.75) (63, 1.72) (64, 1.71) (65, 1.71) (66, 1.68) (67, 1.63) (68, 1.61) (69, 1.61) (70, 1.6) (71, 1.6) (72, 1.59) (73, 1.58) (74, 1.58) (75, 1.58) (76, 1.57) (77, 1.56) (78, 1.55) (79, 1.55) (80, 1.52) (81, 1.51) (82, 1.5) (83, 1.48) (84, 1.47) (85, 1.47) (86, 1.46) (87, 1.44) (88, 1.42) (89, 1.37) (90, 1.36) (91, 1.36) (92, 1.35) (93, 1.32) (94, 1.3) (95, 1.28) (96, 1.26) (97, 1.26) (98, 1.26) (99, 1.24)};
\legend{FLOPS,DF-FLOPS}
\end{axis}
\end{tikzpicture}
    \vspace{-0.8em}
    \caption{ \small Top 100 DF\% in representations produced by SPLADE-Doc with FLOPS and  DF-FLOPS on a sample of 100K passages.
    \vspace{-0.2cm}
    }
    
    \label{fig:stopwrods}
    \Description{Stop words distribution }
\end{figure}

\textbf{Training Details:}  Following~\cite{spladev2}, we train SPLADE-Doc models using the $DistilBERT_{base}$ with a learning rate of $2e^{-5}$ and a batch size of 128. All models are trained for 50K steps without early stopping. The regularization factor $\lambda$ is quadratically increased until 30K steps, with the peak $\lambda$ values for FLOPS models ranging from $10^{-3}$ to $1$ as we found diminishing performance with higher values. The DF-FLOPS models use a higher $\lambda$ value, typically ranging from $10^{-1}$ to $10^{3}$, since DF-FLOPS loss is always less than or equal to FLOPS.
Each training query includes in-batch negatives and 7 hard negatives each iteration (sampled with BM25 following \cite{DBLP:conf/ecir/NguyenMY23}). All of our models were trained using 4 H100 GPUs with PyTorch-Lightning~\cite{pytorch_lightning} and HuggingFace transformers~\cite{huggingface}. Based on a pilot study on MS-Marco dev data, we use $\alpha=0.1$ and $\beta=10$ for our experiments with DF-FLOPS, though we recognize that further ablation of these parameters may yield stronger results.

\textbf{Measuring Efficiency:} In line with our interest in applying LSR in production systems,~\footnote{ In line with our goals of productionization, we used a fully-featured enterprise system (supporting functionality like on-disk indexing, incremental indexing, filtering, etc.), rather than systems oriented for academic testing that do not have these features, such as PISA~\cite{MSMS2019}, SEISMIC~\cite{DBLP:conf/sigir/BruchNRV24}, or BMP~\cite{DBLP:conf/sigir/MalliaST24}. We expect reductions in DFs to improve the efficiency of all these systems, but leave their exploration to future work.}
we perform indexing and retrieval using the open-source Apache Solr\footnote{\url{https://solr.apache.org}} v9 engine (2 cores, 40G RAM, 400G storage).
We index the MS-Marco passage corpus for each model and benchmark the latency over the 6980 queries in the dev set. For each query we measure the retrieval time for the top 10 documents. The retrieval latency includes the query encoding time (with a Bag-of-Words encoder) plus the matching and scoring time. Each measurement is repeated 3 times to calculate the average and 99th percentile latency robustly.

\textbf{Sparsity Metrics: } to measure sparsity, we report the average number of matches in millions~$(M)$ for each query. These are the number of passages Solr considers for a given query (because they share at least a token) to get the top 10 results. The larger the number of matches, the higher the end-to-end retrieval time, as more documents need to be scored and ranked. Second, we report the DF\% of the most frequent token. We found this to be a reliable heuristic for latency in pilot studies; it positively correlates with the number of matches, which drives latency. The top token is often a stopword such as \texttt{"is"} or \texttt{"the"}, but it varies depending on the underlying model. Third, we report the average number of active tokens in a passage representation. For BM25, as there is no expansion, the number of active tokens will be low. For SPLADE family models, document tokens might get dropped or expanded upon to include related tokens by the masked language modeling head. This results in representations denser than BM25's, i.e. with a higher average number of tokens.

\setlength{\tabcolsep}{5pt}
\begin{table*}[t]
\newcolumntype{P}[1]{>{\centering\arraybackslash}p{#1}}

\resizebox{1\textwidth}{!}{%
\begin{tabular}{@{}llcc|cc|cc|  *{6}{P{1cm}}@{}}
& & \multicolumn{2}{c}{\textbf{\small MS-Marco}} & \multicolumn{2}{c}{\textbf{\small TREC 19}} & \multicolumn{2}{c}{\textbf{\small TREC 20}} &
  \multirow{2}{5em}{\textbf{ \small Latency Avg~(ms)}} &
  \multirow{2}{5em}{\textbf{ \small Latency P99~(ms)}} &
  \multirow{2}{5em}{\textbf{ \small Matches Avg~(M)}} &
  \multirow{2}{5em}{\textbf{ \small Top@1 Token DF}} &
  \multirow{2}{5em}{\textbf{ \small Avg. Emb. \\ \ Length}}
   \\
\textbf{\small ID} & \textbf{\small Model Name} & \textbf{\small MRR@10} & \textbf{\small R@1K} & \textbf{\small NDCG@10} & \textbf{\small R@1K} & \textbf{\small NDCG@10} & \textbf{\small R@1K} \\ \midrule
1 & BM25                                & $18.4^*$      & $85.3^*$     & 56.5* & 74.5*  & 47.9* & \textbf{80.5*} & 68.9  & 241.3  & \textbf{0.952}   & 20.6\%  & 27.7 \\ \midrule
2 & SPLADE-Doc w/ \textbf{FLOPS}        & \textbf{32.2} & 92.4         & \textbf{65.6} & 69.6 & \textbf{62.9} & 75.2 & 922.0 & 1945.6 & 8.628 & 95.8\%    &  583.8 \\
3 & \textit{+ Pruning@150}             & 32.0          & 92.1         & 64.6 & 68.9 & 61.8 & 74.0 & 792.1 & 1664.4 & 8.621 & 95.7\%    & 147.6 \\
4 & \textit{+ ↑ $\lambda=0.1$}  & 29.2         & 88.8         & 60.3 & 66.8 & 56.4 & 72.7 & 331.6 & 708.8  & 4.111 &  43.4\%  & 87.6 \\
5 & \textit{+ ↑  $\lambda=1$}    & 28.3         & 88.4         & 56.8 & 67.1 & 53.2 & 72.3 & 160.9 & 347.0  & 1.970 & 17.7\%    & \textbf{33.0} \\ \midrule
6 & SPLADE-Doc w/ \textbf{DF-FLOPS}         & 30.0          & 92.9         & 59.5 & 72.5 & 59.6 & 78.7 & 161.0    & 341.7   & 1.907 & 8.0\%    & 301.6 \\
7 & \textit{+ Pruning@150}            & 29.7         & \textbf{93.0}& 60.2  & \textbf{72.6} & 60.2 & 80.0 & \textbf{87.8}  & \textbf{187.8}  & 1.078 & \textbf{5.2\%}    & 140.3\\ \bottomrule
\end{tabular}
}
\caption{ \small In-domain effectiveness results. On the left, model performance on MS-Marco dev set, TREC'19 and 20. On the right, latency related measures such as average number of matches per query, frequency of the most repeated token in the embedding, and average representation length. A $*$ indicates that the result was copied from~\cite{spladev2} (MS-Marco and TREC 19) or calculated using the PISA Python bindings (TREC 20)~\cite{DBLP:conf/sigir/MacAvaneyM22}.}\vspace{-0.6cm}
\label{tab:results}
\end{table*} 
\newcolumntype{P}[1]{>{\centering\arraybackslash}p{#1}}
\setlength{\tabcolsep}{2pt}
\begin{table}[b]
\resizebox{0.47\textwidth}{!}{%
\begin{tabular}{l*{7}{P{0.8cm}}@{}}
\toprule
 \multirow{2}{5em}{\textbf{BEIR \\Dataset}} & \textbf{BM25} & \multicolumn{4}{c}{\textbf{FLOPS}} & \multicolumn{2}{c}{\textbf{DF-FLOPS}} \\ \cmidrule(lr){3-6} \cmidrule(lr){7-8}
  &  & Base & \textit{Top150} & \textit{$\lambda=.1$} & \textit{$\lambda=1$} & Base & \textit{Top150} \\ \midrule
\textbf{arguana} & $31.5^*$ & 11.16 & 19.06 & 32.61 & 28.83 & 33.25 & \textbf{39.74} \\
\textbf{climate-fever} & $\textbf{21.3}^*$ & 6.89 & 9.15 & 12.49 & 11.96 & 13.44 & 13.58 \\
\textbf{dbpedia-entity} & $27.3^*$ & 31.21 & 30.83 & 30.31 & 30.55 & \textbf{32.73} & 32.26 \\
\textbf{fever} & $\textbf{75.3}^*$ & 57.67 & 50.47 & 58.22 & 60.49 & 63.12 & 60.67 \\
\textbf{fiqa} & $23.6^*$ & 19.64 & 19.28 & 21.29 & 21.08 & \textbf{25.56} & 24.86 \\
\textbf{hotpotqa} & $\textbf{60.3}^*$ & 42.08 & 43.89 & 48.9 & 48.59 & 55.34 & 55.85 \\
\textbf{nfcorpus} & $\textbf{32.5}^*$ & 29.74 & 28.35 & 30.91 & 30.09 & 30.59 & 30.31 \\
\textbf{nq} & $32.9^*$ & \textbf{39.82} & 37.89 & 35.02 & 34.24 & 39.05 & 38.1 \\
\textbf{quora} & $\textbf{78.9}^*$ & 7.56 & 8.13 & 17.14 & 12.39 & 48.1 & 48.35 \\
\textbf{scidocs} & $\textbf{15.8}^*$ & 12.67 & 12.01 & 12.87 & 13.49 & 13.79 & 13.52 \\
\textbf{scifact} & $\textbf{66.5}^*$ & 58.75 & 55.36 & 63.4 & 60.87 & 65.6 & 64.25 \\
\textbf{trec-covid} & $\textbf{65.6}^*$ & 56.17 & 50.15 & 54.65 & 51.78 & 57.52 & 58.56 \\
\textbf{webis-touche2020} & $\textbf{36.7}^*$ & 23.28 & 23.38 & 21.79 & 21.45 & 25.97 & 24.56 \\
  \bottomrule 
\end{tabular}%
}
\caption{\small Out-of-domain performance (NDCG@10) on  BEIR datasets. A $*$ indicates that the result was copied from~\cite{spladev2} \vspace{-0.5cm}}
\label{tab:ood-results}
\end{table}
\section{Results}\label{sec:results}
We present our experimental results on retrieval performance and efficiency on MS-Marco Dev, TREC'19 and TREC'20 in Table~\ref{tab:results}. We also compare the out-of-domain performance using our DF-FLOPS regularization on BEIR in Table~\ref{tab:ood-results}.

\textit{\textbf{RQ1: What is the main driver of latency for SPLADE-Doc when trained with FLOPS?}} In row 2 of Table~\ref{tab:results}, we see that the highest MRR@10 ($32.2$) is achieved by SPLADE-Doc (with FLOPS). However, its average latency is more than 13 times higher than BM25. We can also see that the most frequent token appears in more than 95\% of the documents in the corpus, meaning that when this token is present in the query, nearly the entire collection will be matched, scored, and ranked. This hypothesis is confirmed by the fact that the average number of matches per query for the original SPLADE-Doc model is $8.6M$ out of a corpus of $8.8M$. We plot the DF\% of stopwords\footnote{The list of stopwords is from the NLTK~\cite{bird2006nltk} library.} in Figure~\ref{fig:stopwrods} and observe that many terms have high DFs. \citet{DBLP:journals/corr/abs-2110-11540} also reported the same issue with FLOPS regularization.

\textit{\textbf{RQ2: Can we curb the high-frequency token problem with crude top-k pruning?}} A natural solution is to apply top-k pruning ($pruning@k$) to drop these high-frequency tokens as applied in prior works~\cite{DBLP:conf/sigir/MacAvaneyN0TGF20b,lassance2023static}. An underlying assumption here is that even though these tokens are part of the representation, they would have low weights that can be pruned with little impact to the representation. As can be seen in row 3 of  Table~\ref{tab:results} this approach decreases the latency by $130ms$, yet it is still very high. We argue this occurs due to the high frequency tokens having a high weight in most representations. As evidence, we can see that the top frequent token still appears in 95.7\% of the documents. Furthermore, upon visualization of the document expansions in Table~\ref{example}, we see that frequent tokens (such as \texttt{"of"}, \texttt{"the"} and \texttt{"what"}) are either not dropped by the model, or added as highly weighted expansions. 

\textit{\textbf{RQ3: Can high frequency tokens be mitigated with stronger FLOPS regularization?}} The next intuitive solution is to evaluate FLOPS with a stronger regularization scalar. Our tests with the original setting can be found in row 4 and 5 of  Table~\ref{tab:results}. By increasing the regularization factor to $\lambda=0.1$ and $\lambda=1$ (plus $prunning@150$), we are able to obtain sparser representations, with average documents length of $87.61$ and $33.0$ respectively. Furthermore, the DF\% of the most frequent token significantly decreases to $43.4$ and $17.7$. While these changes significantly reduce the latency (almost by $10x$ for $\lambda = 1$), performance drops substantially, with both MRR@10 and Recall@1000 dropping by around 4 points.   

\textit{\textbf{RQ4: Can DF-FLOPS successfully improve inter-passage token sparsity without significantly dropping performance?}} MS-Marco results on Table~\ref{tab:results} show that DF-FLOPS (row 6) obtains similar latencies to the highly regularized FLOPS setting (row 5), while obtaining $+1.7$ in MRR@10 and $+4.5$ in Recall@1000 (on par with the original setting in row 2). In terms of latency, combining DF-FLOPS with post-hoc pruning@150 (row 7) further almost halves the average latency to $87.7ms$ with minimal impact on MRR@10 and Recall@1000. The results on TREC'19 and TREC'20 are consistent with the findings on MS-Marco. Additionally, we tried static DF-FLOPS by pre-computing the DFs before training but it performed considerably worse, failing to sufficiently reduce the high token frequencies (MRR@10=28.8, Top@1 Token DF=51.7\%).

The DF-FLOPS significantly decreases the latency by lowering the DF\% of high frequent tokens (down to 8\% for the most frequent token). Meanwhile, the representations are not sparser than the ones produced by the original setting (row 4 and row 5 of Table~\ref{tab:results}) as the passage representation length is still high (average token length of 301.6 for row 6). A possible explanation is that penalizing high-frequency tokens forces the model to use more relevant yet less frequent tokens. This reduces the model's necessity to rely on frequent tokens for matching during training. In the example shown in Table~\ref{example}, the model trained with DF-FLOPS often adds semantically relevant tokens. It usually drops NLTK's stopwords, but keeps them when they are semantically relevant.

\textit{\textbf{RQ5: How does DF-FLOPS regularization impact the out-of-domain performance?}}
Table~\ref{tab:ood-results} shows that SPLADE-Doc models trained with DF-FLOPS regularization have a better zero-shot out-of-domain performance than SPLADE-Doc models trained only with FLOPS. SPLADE-v2-max performs comparably with BM25 \cite{spladev2} and SPLADE-Doc exhibits worse out-of-domain generalization as compared to SPLADE. Adding DF-FLOPS outperforms FLOPS-based models on 12 out of 13 BEIR datasets. Moreover, we can see that FLOPS models trained under strong regularization regimes outperform models with lower regularization on several BEIR datasets. An explanation for this can be that over-reliance in high frequency tokens can lead to over-fitting, as dataset dependent information is being encoded in their weights.

\vspace{-0.1cm}
\section{Conclusions}
We show that SPLADE-Doc trained with FLOPS has high retrieval latency due to high frequency tokens being assigned high weights in passage representations and that the FLOPS loss term cannot handle this issue without a substantial drop in performance. To address this, we proposed DF-FLOPS, a simple yet effective regularization method for training SPLADE models that penalizes the usage of high frequency tokens, only using them when they are important, and favors representations that produce small number of matches. Our results show that DF-FLOPS improves the latency by \textbf{10$\times$} in comparison to SPLADE-v2-doc-max, making it competitive to BM25, while maintaining performance at Recall@1000 with a small drop in MRR@10. By shifting the focus from document-level to term-level sparsity, DF-FLOPS regularization offers both a new perspective on sparsification techniques and a tangible improvement in LSR latency, with potential implications for real-world deployment. Future work can evaluate the impact of our method to more effective SPLADE model variants \cite{spladev3,zeng2025scalingsparsedenseretrieval}.

\bibliographystyle{ACM-Reference-Format}
\bibliography{biblio}

\end{document}